\documentstyle[12pt]{article} 
\def \beq {\begin{equation}} 
\def \eeq {\end{equation}} 
\def \bes {\begin{eqnarray}} 
\def \ees {\end{eqnarray}} 
\def \ni {\noindent} 
\def \nn {\nonumber}

\def\e{\varepsilon}
\def\kp{k_{\bot}}
\def\vkp{{\mbox{\boldmath{$k$}}}_{\bot}}
\begin{document} 
\thispagestyle{empty} 
\large 
\begin{titlepage} 
\begin{center} 
 {\Large {\bf 
CASIMIR FORCE UNDER THE INFLUENCE OF REAL CONDITIONS
}} 
 
\vskip 1.5cm 
B. GEYER, G. L. KLIMCHITSKAYA\footnote{On leave from 
North-West Polytechnical University, 
St.Petersburg, Russia, and Federal University of Para\'{\i}ba,
Jo\~{a}o Pessoa, Brazil.} and  
 V. M. MOSTEPANENKO\footnote{On leave from 
A.Friedmann Laboratory for Theoretical
Physics, St.Petersburg, Russia, and Federal University of Para\'{\i}ba,
Jo\~{a}o Pessoa, Brazil.}
\\[8mm]

{\it Center of Theoretical Studies and \\
Institute for Theoretical Physics, Leipzig University, \\
Augustusplatz 10/11, 04109, Leipzig, Germany}  
\end{center} 
\vskip 1.5cm 
Condensed title: Casimir force under real conditions

\end{titlepage} 
 
\sloppy 
\setcounter {page}{2} 
\vspace*{2cm} 
\begin{center} 
{\bf Abstract} 
\end{center} 
\large 
{ 
The Casimir force is calculated analytically for configurations
of two parallel plates and a spherical lens (sphere) above a plate
with account of nonzero temperature, finite conductivity of the 
boundary metal and surface roughness. The permittivity of the metal 
is described by the plasma model. It is proved that in case 
of the plasma model the scattering formalism of quantum field 
theory in Matsubara formulation
underlying Lifshitz formula 
is well defined and  no modifications are needed concerning 
the zero-frequency
contribution. The temperature correction to the Casimir force is found
completely with respect to temperature and perturbatively 
(up to the
second order in the relative penetration depth of electromagnetic
zero-point oscillations into the metal) with respect to finite
conductivity. The asymptotics of low and high temperatures are
presented and contributions of longitudinal and perpendicular
modes are determined separately. Serving as an example, aluminium test
bodies are considered showing good agreement between the
obtained analytical results and previously performed numerical
computations. The roughness correction is formally included and 
formulas are given permitting to calculate the Casimir force under
the influence of all relevant factors.
} 
 
\newpage 
 
\section {Introduction} 

Lately considerable progress had been made both in theoretical and
experimental investigation of the Casimir effect. This effect
predicted by H.B.G.~Casimir${}^1$ more than fifty years ago,
consists in the
interaction of two neutral, conducting bodies placed in vacuum close
to each other. The Casimir effect results from the disturbance by
the conducting boundaries of the zero-point electromagnetic
oscillations. It plays an important role in various fields of physics
such as elementary particle theory, condensed matter physics, 
atomic physics,
gravitation and cosmology, and stimulated new investigations in
mathematical physics (see the
monographs${}^{2-5}$). Recently the Casimir effect found 
applications${}^{6-11}$ for obtaining rather strong constraints on 
hypothetical long-range interactions inspired by 
the physics of extra dimensions,
by unified gauge theories, supersymmetry and supergravity.
Furthermore, topical nanoelectromechanical devices were
proposed${}^{12-14}$ which are based on the use of the Casimir force.

In precision experiments on the measurement of the Casimir
force${}^{14-19}$ different influential factors must be accounted for,
such as nonzero temperature, finite conductivity of the boundary metal
and surface roughness. Theoretically, each factor was investigated in
a number of papers (see, e.g., Refs.~20--24 for the influence of
nonzero temperature, Refs.~22,\,25--29 for the role of finite
conductivity and Refs.~30--36 for the surface roughness). The combined
effect of different corrections was discussed in  Refs.~23,\,24,\,35
(for a detailed discussion of this subject see the recent 
review${}^{37}$).

Investigation of the combined effect of nonzero temperature along with
the finite conductivity of the boundary metal proved to be the most
complicated task leading to controversial results. 
The starting point to theoretically describe this effect
is the Lifshitz theory${}^{38}$ originally being developed 
for dielectrics.
In order to describe on the
base of Lifshitz' theory
the Casimir force between plates 
made of an ideal metal a special prescription was suggested in
Ref.~22 which demands to consider first the limit of infinite dielectric
permittivity before setting the frequency equal to zero.
Using this prescription, the results of the Lifshitz theory 
for ideal metals
agreed with the results obtained by the application of quantum field
theory using the idealized boundary conditions.${}^{20,21}$

In the last year, several authors attempted 
to apply the Lifshitz theory to calculate
the Casimir force between plates made of real metals. In Ref.~23 the
plasma model was used to describe the dependence of dielectric
permittivity on frequency. In Ref.~39 the Drude dielectric function
(being a generalization of the plasma model by taking into account the
relaxation processes) was substituted into the Lifshitz formula.
However, in the limit of zero relaxation frequency the results of
Ref.~39  do not coincide with those of Ref.~23 although Drude's model
turns into plasma model in this limit. In Ref.~24 the results of
Ref.~23 obtained using the plasma model were independently confirmed
and doubts were casted on the calculations of Ref.~39  using the Drude
model. It was noticed${}^{24}$  that the high temperature
asymptotics of the Casimir force between real metals computed
in Ref.~39 is two times smaller than in the case of an ideal metal ---
independently of how high the conductivity of the real metal is --- which
is a nonphysical property.  As was noted in Ref.~40, the computations of
Ref.~39 are also in contradiction with the experiment.${}^{15}$ 

To improve this situation in  Ref.~41 a detailed computation of the
Casimir force at nonzero temperature were performed based on Drude's
model and a Lifshitz formula with some modified zero-frequency term.
The modification made is based on a generalization of the prescription
of Ref.~22 for the case of real metals. The results of Ref.~41 are
in agreement with both limiting cases of an ideal metal and a metal
described by the plasma model. In Refs.~42,\,43 one more result was
obtained for the temperature Casimir force between real metals which,
however,
disagrees with both the results of Ref.~39 from one side and of
Refs.~23,\,24 from the other. According to Refs.~42,\,43 at small
frequencies all real metals are indistinguishable from the ideal metal.
This leads to the absence of any finite conductivity correction to
the Casimir force starting from moderate separations of several 
micrometers between the test bodies. Even more, in the approach of
Refs.~42,\,43 this property is independent on the quality of the real
metal.

As is seen from the above, at the time being there is 
no agreement in the literature
concerning the calculation of the Casimir force acting between real metals
at nonzero temperature. Different results are obtained on this subject
by different authors starting from one and the same theoretical
foundation given by the Lifshitz formula for dielectrics. In the present
paper we discuss the scattering formalism of quantum field theory
at finite temperature
in the Matsubara formulation underlying Lifshitz formula
and we argue that it leads to
well defined and consistent results both physically and
mathematically if the dielectric function is
described by the plasma model.
This gives additional support to the results of Refs.~23,\,24.
Contrary,
when the dielectric function of the Drude model is used the scattering
formalism becomes inconsistent causing the nonphysical results of
Refs.~39,\,42,\,43. In this case the modification of Lifshitz formula
suggested in Ref.~41 is needed. 

Below, starting from the dielectric function
of the plasma model we calculate the Casimir force for the configurations
of two parallel plates and a spherical lens (sphere) above a plate
made of real metals. The temperature corrections are taken into account
completely and the effects of finite conductivity of the boundary metal
are treated perturbatively up to second order in some small parameter
having the meaning of the relative penetration depth of electromagnetic
zero-point oscillations into the metal. Note that in the analytical
computations of the previous paper${}^{23}$ the finite conductivity
corrections were calculated up to the first order only, whereas in
Ref.~24 both corrections 
(due to nonzero temperature and finite conductivity)
were treated perturbatively. The analytical results obtained below are
compared with the results of numerical computations and good agreement
is observed for all space separations exceeding the plasma wavelength
of the boundary metal. We also include roughness correction and
demonstrate a way how to take into account the influence of real
conditions (which include all three types of corrections) onto the
Casimir force.

The paper is organized as follows. In Sec.2 the general formalism is
presented and the scopes of its consistency are elucidated. In Sec.3
the temperature correction to the Casimir force is calculated for the 
configuration of two parallel plates up to the second perturbation
order in the relative penetration depth. Sec.4 contains the analogous
results for the configuration of a sphere (spherical lens) above a plate.
In Sec.5 the roughness correction is taken into account along with
nonzero temperature and finite conductivity corrections. In Sec.6 
the reader finds our conclusions and discussion.

\section{General formalism and its scopes}

Lifshitz' original derivation${}^{38}$ of the Casimir force at
nonzero temperature acting between two dielectric semispaces
separated by a gap was based on the assumption that the dielectric 
materials can be considered as continuous media characterized by 
randomly fluctuating sources. The modern derivation${}^{37}$ is
based on quantum field theory at nonzero temperature, $T\neq 0$,
in the Matsubara formulation. 
Thereby one considers
the Euclidean version with the electromagnetic field periodic in the
Euclidean time variable within the time interval 
$\beta=\hbar/(k_B T)$, where $k_B$ and $\hbar$ are the Boltzmann and 
Planck constants, respectively.

Let us consider two dielectric semispaces with frequency dependent
permittivity $\e (\omega)$ restricted by two planes at
$z=\pm a/2$ and separated by a vacuum gap of width $a$ between them.
Due to periodicity in the time coordinate, the frequency spectrum is
discrete $\omega_l=2\pi l/\beta$, whith
$l=...\,-2,\,-1,\,0,\,1,\,2,\,...\,$.
The calculation of the free energy is reduced to the solution of a
one-dimensional scattering problem on the $z$-axis. In fact an
electromagnetic wave which is coming from the left, or from the right,
in the dielectric semispaces will be scattered on the vacuum gap
and there will be a reflected and a transmitted wave. The free energy
of the field per unit area, $E_{ss}$, is calculated with the help
of $\zeta$-regularization method. The result is${}^{37}$ 
\beq
E_{ss}(a)=-\frac{\hbar}{2\beta}
\sum\limits_{l}\int
\frac{d\vkp}{(2\pi)^2}\left[
\ln s_{11}^{||}\left(i\xi_l,\vkp
\right)+
\ln s_{11}^{\bot}\left(i\xi_l,\vkp
\right)\right],
\label{1}
\eeq
\ni
where $s_{11}^{||}\left(i\xi_l,\vkp\right)$ and
 $s_{11}^{\bot }\left(i\xi_l,\vkp\right)$ 
are the scattering coefficients for parallel and
perpendicular polarizations, respectively,
$\vkp =(k_x,k_y)$ is the wave vector
in the planes perpendicular to the $z$-axis,
and $\xi_l=2\pi l/\beta$. The solution of the
scattering problem reads${}^{37}$ 
\bes
&&
s_{11}^{||}=
\frac{4\e(i\xi_l)k_lq_le^{k_la}}{\left[\e(i\xi_l)q_l+
k_l\right]^2e^{q_la}-\left[\e(i\xi_l)q_l-k_l\right]^2e^{-q_la}},
\nn \\
&&
s_{11}^{\bot}=
\frac{4k_lq_le^{k_la}}{\left(q_l+
k_l\right)^2e^{q_la}-\left(q_l-k_l\right)^2e^{-q_la}},
\label{2}
\ees
\ni
where the following notations are introduced
\beq
q_l=\sqrt{\frac{\xi_l^2}{c^2}+{\kp}^2}, \quad
k_l=\sqrt{\e (i\xi_l)\frac{\xi_l^2}{c^2}+{\kp}^2},
\quad \kp\equiv|\vkp |.
\label{3}
\eeq
\ni
Now we substitute (\ref{2}) into (\ref{1}) and perform renormalization
in order to get 
the free energy equal to zero in the case
of infinitely far remote plates.${}^{28}$ 
In terms of reflection coefficients
$r_{||},\> r_{\bot}$ for the electromagnetic waves of the two different
polarizations one obtains
\bes
&&
E_{ss}(a)=\frac{k_B T}{4\pi}
\sum\limits_{l}
\int\limits_{0}^{\infty}
\kp d\kp
\left\{
\ln\left[1- r_{||}^2\left(\xi_l,\kp\right)
e^{-2aq_l}\right]\right.
\nn \\
&&\phantom{aaaaaaaaaaaaaaaaaa}
\left.+
\ln\left[1- r_{\bot}^2\left(\xi_l,\kp\right)
e^{-2aq_l}\right]
\right\},
\label{4}
\ees
\ni
where  
\beq
r_{||}^2(\xi_l,\kp)=\left(
\frac{\e (i\xi_l)q_l-k_l}{\e (i\xi_l)q_l+k_l}\right)^2,
\qquad
r_{\bot}^2(\xi_l,\kp)=\left(\frac{q_l-k_l}{q_l+k_l}\right)^2.
\label{5}
\eeq

The Casimir force per unit area acting between two semispaces is
obtained as $-\partial E_{ss}/\partial a$ with the result
\bes
&&
F_{ss}(a)=-\frac{k_B T}{2\pi}
\sum\limits_{l}
\int\limits_{0}^{\infty}
\kp d\kp q_l
\left\{
\left[r_{||}^{-2}\left(\xi_l,\kp\right)
e^{2aq_l}-1\right]^{-1}\right.
\nn \\
&&\phantom{aaaaaaaaaaaaaaaaaa}
\left.+
\left[r_{\bot}^{-2}\left(\xi_l,\kp\right)
e^{2aq_l}-1\right]^{-1}
\right\}.
\label{6}
\ees
\ni
This equation, up to change of variables, coincides with the original
Lifshitz expression${}^{38}$ for the Casimir force between dielectrics
at nonzero temperature.

Our aim is to apply Eqs.~(\ref{4})--(\ref{6}) in the case of test
bodies made of real metals. Then, the zero-frequency contribution
for $\xi_0=0$ in Eq.~(\ref{4}) may become indefinite 
 in the case of perpendicular polarization
when the dielectric permittivity turns into infinity. 
For example, let us
consider the Drude dielectric function on the imaginary axis
\beq
\e (i\xi)=1+\frac{\omega_p^2}{\xi(\xi+\gamma)},
\label{7}
\eeq
\ni
where $\omega_p$ is the plasma frequency, $\gamma$ is the relaxation
frequency, which gives a good approximation of dielectric properties
for some metals, e.g., for aluminium. 
Note that for dielectrics $\e (i\xi)\to \e_0$ when $\xi\to 0$. 
It is evident that
\beq
\xi^2\e (i\xi)\to 0 \quad\mbox{when}{\ } \xi\to 0,
\label{8}
\eeq
\ni
both for Drude's model and dielectrics.
By this reason, it follows from Eq.~(\ref{3}) that 
$q_0=k_0=\kp$. Strictly speaking, the mathematical derivation leading
to Eq.~(\ref{2}) for $s_{11}^{\bot}$ is inapplicable in that case.
Instead, the direct solution of the scattering problem 
in the case $q_0=k_0$
gives the result that $s_{11}^{\bot}$ is arbitrary
and $s_{12}^{\bot}=0$,
where $s_{12}$ is the nondiagonal element of the scattering matrix.
In the case of dielectrics the unitarity condition is valid
which immediately leads to $|s_{11}^{\bot}|^2=1$ and, due to
dispersion relation, to $s_{11}^{\bot}=1$. In fact, the same result is
obtained from Eq.~(\ref{2}) in the limit $q_0=k_0$.
However, as to the case of the Drude model, 
which describes a medium with
dissipation, the unitarity condition is not applicable, and, therefore,
the scattering coefficient $s_{11}^{\bot}$ remains indefinite.
Because of this, the direct application of Lifshitz formula (as in
Ref.~39) leads to incorrect results. To apply the Lifshitz formula
at nonzero temperature
in combination with Drude model some additional prescription is
needed to give the definite value to the zero-frequency term
(see Ref.~41 for details).

 At the same
time for the longitudinal polarization the scattering coefficient
is well defined in the limit of zero frequency. In the case of the Drude 
model, up to terms independent of $a$, it has the limiting
value
\beq
s_{11}^{||}\to \left(1-e^{-2a\kp}\right)^{-1}
\quad\mbox{when} {\ }\xi\to 0,
\label{9}
\eeq
\ni
i.e., the same as for ideal metals. For dielectrics
\beq
s_{11}^{||}\to \left[\left(\frac{\e_0+1}{\e_0-1}\right)^2
-e^{-2a\kp}\right]^{-1}
\quad\mbox{when} {\ }\xi\to 0.
\label{10}
\eeq
\ni
Thus, the transition from Eq.~(\ref{1})
to Eqs.~(\ref{4}), (\ref{6}) is unjustified in the case of Drude's
dielectric function. This explains why the nonphysical results were
obtained when substituting Eq.~(\ref{7}) into Eq.~(\ref{6})
(see Introduction). 

Another model of the dielectric function is the plasma one,
\beq
\e (i\xi)=1+\frac{\omega_p^2}{\xi^2},
\label{11}
\eeq
\ni
which is the limiting case of (\ref{7}) when the relaxation frequency
$\gamma$ goes to zero. In the case of the plasma dielectric function
\beq
\xi^2\e (i\xi)\to \omega_p^2\neq 0 \quad\mbox{when}{\ } \xi\to 0.
\label{12}
\eeq
\ni
As a consequence, here $q_0\neq k_0$ and
the limiting value of the perpendicular scattering
coefficient is given by
\beq
s_{11}^{\bot}\to\left[
\frac{\kp +\sqrt{\frac{\omega_p^2}{c^2}+{\kp}^2}}{\kp-
\sqrt{\frac{\omega_p^2}{c^2}+{\kp}^2}} 
-e^{-2a\kp}\right]^{-1}
\quad\mbox{when}{\ } \xi\to 0.
\label{13}
\eeq
\ni
The limiting value of the longitudinal scattering coefficient in
the case of plasma model is the same as in Eq.~(\ref{9}). Because of
this, the scattering problem is well defined for the dielectric
function (\ref{11}) and Eqs.~(\ref{4}) and (\ref{6}) can be reliably
applied to calculate the free energy and the Casimir force.
No additional prescriptions or modifications are admissible
in the case of plasma model.
 Because of this, the manipulations of Refs.~42,\,43
changing the zero-frequency term of Lifshitz formula for both
plasma and Drude models seem to be unfounded. In the case of the plasma
model they lead to disagreement with the results of Refs.~23,\,24
where no modifications of Lifshitz formula have been made. 

In the next
section the analytical computations of the temperature correction
to the Casimir force are performed on the basis of Eqs.~(\ref{4}), 
(\ref{6}) by using the plasma model (\ref{11}).

\section{Temperature correction to the Casimir force for two
parallel plates made of real metal}

We start with the Lifshitz formula (\ref{6}) and rewrite it in terms
of dimensionless variables
\beq
y=2aq_l=2a\sqrt{\frac{\xi_l^2}{c^2}+{\kp}^2}, \qquad
x_l=2a\frac{\xi_l}{c}
\label{14}
\eeq
\ni
resulting in
\bes
&&
F_{ss}(a)=-\frac{k_B T}{16\pi a^3}
\sum\limits_{l}
\int\limits_{|x_l|}^{\infty}
y^2 dy
\left\{
\left[r_{||}^{-2}\left(x_l,y\right)
e^{y}-1\right]^{-1}\right.
\nn \\
&&\phantom{aaaaaaaaaaaaaaaaaa}
\left.+
\left[r_{\bot}^{-2}\left(x_l,y\right)
e^{y}-1\right]^{-1}
\right\}.
\label{15}
\ees
\ni
Here the reflection coefficients in new variables take the form
\beq
r_{||}(x_l,y)=
\frac{\e y-\sqrt{(\e -1)x_l^2+y^2}}{\e y+\sqrt{(\e -1)x_l^2+y^2}},
\quad
r_{\bot}(x_l,y)=
\frac{y-\sqrt{(\e -1)x_l^2+y^2}}{y+\sqrt{(\e -1)x_l^2+y^2}},
\label{16}
\eeq
\ni
and $\e\equiv\e\left[icx_l/(2a)\right]$.

To separate within (\ref{15}) the contribution of temperature
$T=0$
and the temperature correction one can use the representation
of this formula in terms of continuous $x$ instead of discrete
summation in $x_l$. Applying the Poisson summation formula one
obtains from (\ref{15})${}^{20,22,23}$
\bes
&&
F_{ss}(a)=-\frac{\hbar c}{32\pi^2 a^4}
\sum\limits_{l}
\int\limits_{0}^{\infty}
y^2 dy
\int\limits_{0}^{y}
dx \cos(ltx)
\nn \\
&&\phantom{aaaaaa}
\times\left\{
\left[r_{||}^{-2}\left(x,y\right)
e^{y}-1\right]^{-1}+
\left[r_{\bot}^{-2}\left(x,y\right)
e^{y}-1\right]^{-1}
\right\},
\label{17}
\ees
\ni
where $t=T_{eff}/T$, $k_BT_{eff}\equiv\hbar c/(2a)$. 
The reflection coefficients preserve their form (\ref{16})
with a change $x_l\to x$.

In Eq.~(\ref{17}) the term with $l=0$ is the Casimir force at
zero temperature,${}^{22}$ whereas the terms with $l\neq 0$ 
represent the temperature corrections. For further needs it is
convenient to write the temperature correction of the Casimir force
acting between real metals as a sum of  longitudinal and
perpendicular contributions,
\beq
\Delta_T F_{ss}(a)=
\Delta_T^{||} F_{ss}(a)+
\Delta_T^{\bot} F_{ss}(a),
\label{18}
\eeq
\ni
where
\bes
&&
\Delta_T^{||(\bot)} F_{ss}(a)
=-\frac{\hbar c}{32\pi^2 a^4}
\sum\limits_{l=1}^{\infty}
\int\limits_{0}^{\infty}
y^2 dy
\int\limits_{0}^{y}
dx \cos(ltx)
\nn \\
&&\phantom{aaaaaaa}
\times\left[r_{||(\bot)}^{-2}\left(x,y\right)
e^{y}-1\right]^{-1}.
\label{19}
\ees

To compute the temperature correction (\ref{18}), (\ref{19}) one
should use some specific
functional dependence of the dielectric permittivity
on the frequency. Here we use the plasma model (\ref{11}) for which
the theory is well defined (see Sec.2). In terms of dimensionless
variables the dielectric function (\ref{11}) is given by
\beq
\e =\e (x)=1+\frac{{\tilde\omega}_p^2}{x^2}=
 1+\frac{4a^2}{\delta_0^2x^2},
\label{20}
\eeq
\ni
where $ {\tilde\omega}_p=2a\omega_p/c$, $\delta_0=c/\omega_p$
is the effective penetration depth of electromagnetic zero-point 
oscillations
into the real metal. For the space separations between the plates
$a\gg\delta_0$ the natural small parameter is $\delta_0/a\ll 1$.
In fact this condition is valid for $a>\lambda_p$, where
$\lambda_p=2\pi c/\omega_p$ is the effective plasma wavelength,
since $\delta_0=\lambda_p/(2\pi)$. 

Here we calculate the temperature correction (\ref{18}), (\ref{19})
analytically taking completely into account the nonzero temperature and
using the perturbation theory up to the second power in the small
parameter  $\delta_0/a$ in order to take approximate account
of the finite conductivity of the boundary metal. To perform the
computations let us expand first the expressions in Eq.~(\ref{19}),
containing reflection coefficients, in powers of $\delta_0/a$.
The result is
\bes
&&
\left[r_{||}^{-2}\left(x,y\right)
e^{y}-1\right]^{-1}=
\frac{1}{e^y-1}-2\frac{\delta_0}{a}\frac{x^2e^y}{y(e^y-1)^2}
\nn \\
&&
\phantom{aaaaaa}
+2\left(\frac{\delta_0}{a}\right)^2
\frac{x^4e^y(e^y+1)}{y^2(e^y-1)^3}+
O\left(\frac{\delta_0^3}{a^3}\right),
\label{21} \\
&&
\left[r_{\bot}^{-2}\left(x,y\right)
e^{y}-1\right]^{-1}=
\frac{1}{e^y-1}-2\frac{\delta_0}{a}\frac{ye^y}{(e^y-1)^2}
\nn \\
&&
\phantom{aaaaaa}
+2\left(\frac{\delta_0}{a}\right)^2
\frac{y^2e^y(e^y+1)}{(e^y-1)^3}+O\left(\frac{\delta_0^3}{a^3}\right).
\nn
\ees

Substituting (\ref{21}) into (\ref{19}) and calculating integrals
with the help of the formulas 3.951(12,\,13) of Ref.~44 
one obtains finally the contribution of
the longitudinal mode
\bes
&&
\Delta_T^{||} F_{ss}(a)
=-\frac{\hbar c}{16\pi^2 a^4}
\sum\limits_{l=1}^{\infty}
\left\{\frac{1}{(lt)^4}-
\frac{\pi^3}{lt}\frac{\coth(\pi lt)}{\sinh^2(\pi lt)}
\right.
\nn \\
&&\phantom{aaa}
+2\frac{\delta_0}{a}\left[\frac{\pi}{(lt)^3}\coth(\pi lt)+
\frac{\pi^2}{(lt)^2\sinh^2(\pi lt)}
\left(\vphantom{coth^2(tl)}
1+\pi lt\coth(\pi lt)
\right.\right.
\label{22}\\
&&\phantom{aaaaaaaa}
\left.\left.
+(\pi lt)^2-3(\pi lt)^2-\coth^2(\pi lt)\right)\right]
\nn \\
&&\phantom{aaa}
+2\left(\frac{\delta_0}{a}\right)^2\frac{\pi^3}{\sinh^2(\pi lt)}
\left[-4\pi +12\pi \coth^2(\pi lt)+ 
7\pi^2 lt\coth(\pi lt)
\right.
\nn \\
&&\phantom{aaaaaaaaaa}
\left.\left.
-12\pi^2 lt \coth^3(\pi lt)\right]
\vphantom{\frac{\pi^3}{\sinh^2(\pi lt)}}
\right\}.
\nn
\ees
\ni
Quite analogically the contribution of the perpendicular mode is
\bes
&&
\Delta_T^{\bot} F_{ss}(a)
=-\frac{\hbar c}{16\pi^2 a^4}
\sum\limits_{l=1}^{\infty}
\left\{\frac{1}{(lt)^4}-
\frac{\pi^3}{lt}\frac{\coth(\pi lt)}{\sinh^2(\pi lt)}
\right.
\nn \\
&&\phantom{aaa}
+2\frac{\delta_0}{a}\frac{\pi^3}{lt\sinh^2(\pi lt)}
\left[3\coth(\pi lt)+\pi lt-
3\pi lt\coth^2(\pi lt)
\right]
\nn \\
&&\phantom{aaa}
+8\left(\frac{\delta_0}{a}\right)^2\frac{\pi^3}{lt\sinh^2(\pi lt)}
\left[-3\coth(\pi lt)+2(\pi lt)^2\coth^2(\pi lt)
\right.
\label{23} \\
&&\phantom{aaaaaaaaaa}
\left.\left.
-2\pi lt +6\pi lt\coth^2(\pi lt)
-3(\pi lt)^2 \coth^3(\pi lt)\right]
\vphantom{\frac{\pi^3}{\sinh^2(\pi lt)}}
\right\}.
\nn
\ees
\ni
Finally, the total temperature correction (\ref{18}) is given by
\bes
&&
\Delta_T F_{ss}(a)
=-\frac{\hbar c}{8\pi^2 a^4}
\sum\limits_{l=1}^{\infty}
\left\{\frac{1}{(lt)^4}-
\frac{\pi^3}{lt}\frac{\coth(\pi lt)}{\sinh^2(\pi lt)}
\right.
\nn \\
&&\phantom{aaa}
+\frac{\delta_0}{a}\frac{\pi^3}{lt\sinh^2(\pi lt)}
\left[\frac{1}{(\pi lt)^2}\sinh(\pi lt)\cosh(\pi lt)+
4\coth(\pi lt)
\right.
\nn \\
&&\phantom{aaaaaaaaaaaaaa}
\left.
+2\pi lt-
6\pi lt\coth^2(\pi lt)+\frac{1}{\pi lt}\
\right]
\label{24} \\
&&\phantom{aaa}
+3\left(\frac{\delta_0}{a}\right)^2\frac{\pi^3}{lt\sinh^2(\pi lt)}
\left[-4\pi lt +5(\pi lt)^2\coth(\pi lt)+12\pi lt\coth^2(\pi lt)
\right.
\nn \\
&&\phantom{aaaaaaaaaa}
\left.\left.
-8(\pi lt)^2 \coth^3(\pi lt)
-4\coth(\pi lt)\right]
\vphantom{\frac{\pi^3}{\sinh^2(\pi lt)}}
\right\}.
\nn
\ees
\ni
Note that the first two terms on the right-hand side of (\ref{24})
which are of zeroth order in $\delta_0/a$ coincide with the well-known
result for the ideal metal${}^{21,22}$, whereas the coefficient of
the first power in $\delta_0/a$ was first obtained in Ref.~23.

In the case of low temperatures (small separations), $t\gg 1$, one
can substitute the hyperbolic functions by their asymptotics. 
Preserving the largest of the
exponentially small contributions and performing summations of
the power ones we obtain from (\ref{24})
\bes
&&
\Delta_T F_{ss}(a)\approx
-\frac{\hbar c}{8\pi^2 a^4}
\left\{\frac{\pi^4}{90t^4}-\frac{4\pi^3}{t}e^{-2\pi t}
\right.
\label{25} \\
&&
\phantom{aaaaa}
\left.
+\frac{\delta_0}{a}\left[
\frac{\pi}{t^3}\zeta(3)-16\pi^4e^{-2\pi t}\right]
-36\pi^5t\left(\frac{\delta_0}{a}\right)^2
e^{-2\pi t}\right\},
\nn
\ees
\ni
where $\zeta(z)$ is Riemann's zeta function. It is seen that the
second perturbative order in $\delta_0/a$ is exponentially small in
$t$ and does not contain purely powers in $t$ contributions like the
first order term of (\ref{25}). This is in agreement with the
perturbation results of Ref.~24 where the double perturbation
theory in the small parameters $\delta_0/a$  and $1/t$ was developed.
At the same time, under the natural supposition 
$(\delta_0/a)t\sim 1$ the second order term turns out to be 
approximately 90 times larger than the exponentially small
contribution in the zeroth order term and 7 times larger 
than the exponentially small contribution in the first order term.

Now consider the case of high temperatures (large separations) when
$t\ll 1$. It is more simple to extract it not from Eq.~(\ref{24}) but
directly
from Eqs.~(\ref{18}), (\ref{19}), (\ref{21}). To do this one should
perform the integration with respect to $x$ in the same way as above
and then change the order of summation and integrations with respect
to $y$. Due to the smallness of $t$ all summations can be performed
by the use of the formula${}^{44}$
\beq
\sum\limits_{l=0}^{\infty}
\frac{\sin(lty)}{l}= \frac{\pi-ty}{2},
\label{26}
\eeq
\ni
which is valid for $0<ty<2\pi$. In the further integrations with respect
to $y$ all functions under the integrals decrease 
with $y$ as $\exp(-y)$,
so that the infinite upper limit of the integration 
can be changed for
$\tilde{y}=(2\pi/t)-\alpha$, where $\alpha>0$, with the required accuracy.
As a result, the high temperature limit of the temperature correction
to the Casimir force between real metals is given by
\bes
&&
\Delta_T F_{ss}(a)\approx
-\frac{\hbar c}{8\pi^2 a^4}
\left\{\frac{\pi\zeta(3)}{t}-\frac{\pi^4}{30}
\right.
\label{27} \\
&&
\phantom{aaaaa}
\left.
+\frac{\delta_0}{a}\left[
-\frac{3\pi}{t}\zeta(3)+\frac{8\pi^4}{45}\right]
+\left(\frac{\delta_0}{a}\right)^2
\left[\frac{12\pi}{t}\zeta(3)-\frac{4\pi^4}{5}\right]
\right\}.
\nn
\ees

Let us discuss the application range of the analytical result (\ref{24})
and the asymptotic representations  (\ref{25}) and (\ref{27}). 
Bearing in mind that according to Eq.~(\ref{17}) the Casimir force
at nonzero temperature and finite conductivity is given by
\beq
F_{ss}(a)=F_{ss}(a;T=0)+\Delta_T F_{ss}(a),
\label{28}
\eeq
\ni
it is convenient to compute the quantity
\beq
k_{ss}=\frac{\Delta_T F_{ss}(a)}{F_{ss}(a;T=0)}.
\label{29}
\eeq
\ni
Then the value of $(1+k_{ss})$ has the meaning of a correction factor.
Indeed, multiplying the Casimir force $F_{ss}(a;T=0)$ computed with
account of finite conductivity at zero temperature by $(1+k_{ss})$ 
one obtains the Casimir force at both nonzero temperature and finite
conductivity. Note that $F_{ss}(a;T=0)$ was computed in a number of
papers${}^{27-29}$, and below we use the numerical 
and analytical results obtained
there to calculate $k_{ss}$.

In Table 1 the values of $k_{ss}$ are presented at several 
separations computed by the use of Eq.~(\ref{24}) (second column),
by the use of low-temperature asymptotics (third column), and
high-temperature asymptotics (fourth column). In all computations,
as an example, the value of the plasma frequency
$\omega_p=12.5\,$eV is used as for aluminium${}^{45}$  (this
corresponds to the plasma wavelength of approximately
$\lambda_p\approx 99\,$nm). It is seen from Table 1 that the
asymptotic of low temperatures gives the same values of $k_{ss}$ 
as Eq.~(\ref{24}) at all separations $a\leq 2\,\mu$m (for
$a>3\,\mu$m it is not applicable). Comparing data of columns two
and four one can conclude that the asymptotics of high temperatures
works good for $a\geq 7\,\mu$m  and is not applicable for 
$a<5\,\mu$m. In the transition region 
$3\,\mu\mbox{m}\leq a\leq 5\,\mu$m 
neither of the asymptotics but Eq.~(\ref{24}) itself
should be used to compute the temperature correction to the Casimir
force acting between real metals. It is noticeable, that data of
column 2 are practically the same irrespective of whether one or two
perturbation orders are taken into account. These data coincide also
with the results of numerical computations by 
Eqs.~(\ref{18})--(\ref{20}) in all separation range
$0.1\,\mu\mbox{m}\leq a\leq 10\,\mu$m.

It is interesting also to discuss the
comparative contribution to the temperature correction which results 
from the longitudinal and perpendicular modes given by 
Eqs.~(\ref{22}), (\ref{23}).
At small separations the contribution of
$\Delta_{T}^{||}F_{ss}$ to the temperature correction dominates the
contribution of $\Delta_{T}^{\bot}F_{ss}$. By way of example,
at $a=0.1\,\mu$m one has 
$\Delta_{T}^{||}F_{ss}/\Delta_{T}^{\bot}F_{ss}=43.1$.
This ratio, however, quickly decreases with the increase of space
separation. Thus, at $a=0.3\,\mu$m  it is equal to 5.67, whereas
at $a=0.5\,\mu$m and  $a=0.7\,\mu$m it equals to
 2.68 and 1.86, respectively.
At large separations both modes give almost equivalent contribution
to the temperature correction. For example, at $a=7\,\mu$m 
the abovementioned ratio is equal to 1.015 and at $a=10\,\mu$m 
it is equal to 1.01.

\section{Temperature correction for a sphere above a plate
made of real metal}

In most of the experiments on measurement of the Casimir force the
configuration of a sphere (spherical lens) placed above a plate
(semispace) is used${}^{15-18}$ (in fact the configuration of two
crossed cylinders, as in Ref.~19, is equivalent to it). The
expression for the Casimir force at nonzero temperature acting
in this configuration can be obtained by means of the proximity force
theorem${}^{46}$ 
\beq
F_{sl}(a)=2\pi R E_{ss}(a),
\label{30}
\eeq
\ni
where $E_{ss}(a)$ is the free energy per unit area of the two plates
defined in Eq.~(\ref{4}), $R$ is the curvature radius of the
sphere (spherical lens). In terms of the dimensionless variables 
of Eq.~(\ref{14}) the force
acting between a lens and a plate is
\bes
&&
F_{sl}(a)=\frac{k_B TR}{8 a^2}
\sum\limits_{l}
\int\limits_{|x_l|}^{\infty}
y dy
\left\{
\ln\left[1-r_{||}^{2}\left(x_l,y\right)
e^{-y}\right]\right.
\nn \\
&&\phantom{aaaaaaaaaaaaaaaaaa}
\left.+
\ln\left[1-r_{\bot}^{2}\left(x_l,y\right)
e^{-y}\right]
\right\},
\label{31}
\ees
\ni
where the reflection coefficients $r_{||(\bot)}$ are defined in (\ref{16}).

After applying the Poisson summation formula, Eq.~(\ref{31}) can
be rewritten in the form analogical to Eq.~(\ref{17})
\bes
&&
F_{sl}(a)=\frac{\hbar cR}{16\pi a^3}
\sum\limits_{l}
\int\limits_{0}^{\infty}
y dy
\int\limits_{0}^{y}
dx \cos(ltx)
\nn \\
&&\phantom{aaaaaaa}
\times\left\{
\ln\left[1-r_{||}^{2}\left(x,y\right)
e^{-y}\right]+
\ln\left[1-r_{\bot}^{2}\left(x,y\right)
e^{-y}\right]
\right\}.
\label{32}
\ees
\ni
Once more, the term with $l=0$  is the Casimir force at zero
temperature, the terms with $l\neq 0$ represent the temperature
corrections to it. In accordance with Eq.~(\ref{18}) 
the temperature correction can be splitted 
into a sum of longitudinal
and perpendicular contributions (with a change of index $ss\to sl$)
expressed by
\bes
&&
\Delta_T^{||(\bot)} F_{sl}(a)
=\frac{\hbar cR}{8\pi a^3}
\sum\limits_{l=1}^{\infty}
\int\limits_{0}^{\infty}
y dy
\int\limits_{0}^{y}
dx \cos(ltx)
\nn \\
&&\phantom{aaaaaaa}
\times\ln\left[1-r_{||(\bot)}^{2}\left(x,y\right)
e^{-y}\right].
\label{33}
\ees

To compute the temperature correction
analytically by using the plasma model we substitute Eq.~(\ref{20})
into Eq.~(\ref{33}) and, again, expand into powers of the small
parameter  $\delta_0/a$ 
\bes
&&
\ln\left[1-r_{||}^{2}\left(x,y\right)
e^{-y}\right]=
\ln(1-e^{-y})+2\frac{\delta_0}{a}\frac{x^2}{y(e^y-1)}
\nn \\
&&
\phantom{aaaaaa}
-2\left(\frac{\delta_0}{a}\right)^2
\frac{x^4e^y}{y^2(e^y-1)^2}+
O\left(\frac{\delta_0^3}{a^3}\right),
\label{34} \\
&&
\ln\left[1-r_{\bot}^{2}\left(x,y\right)
e^{-y}\right]=
\ln(1-e^{-y})+2\frac{\delta_0}{a}\frac{y}{e^y-1}
\nn \\
&&
\phantom{aaaaaa}
-2\left(\frac{\delta_0}{a}\right)^2
\frac{y^2e^y}{(e^y-1)^2}+O\left(\frac{\delta_0^3}{a^3}\right).
\nn
\ees

Let us calculate first the longitudinal temperature correction which
is obtained by the substitution of the first equality from (\ref{34})
into (\ref{33}). All integrals with respect to $x$ are trivial. 
The resulting integrals with respect to $y$ can be found in Ref.~44
(formulas 3.951(12,\,13))
except of the following
one which, however, also can be computed analytically:
\bes
&&
\int\limits_{0}^{\infty}
\frac{dy}{y}\frac{e^y}{(e^y-1)^2}\left[\sin(lty)-lty\cos(lty)
\right]
\label{35} \\
&&
=\frac{\pi l^2t^2}{4}\coth(\pi lt)-\frac{1}{4\pi}\left[
\frac{\pi^2}{6}+2\pi lt\ln\left(1-e^{-2\pi lt}\right)-
\frac{2\pi^2 l^2t^2}{e^{2\pi lt}-1}-
{\mbox{Li}}_2\left(e^{-2\pi lt}\right)\right],
\nn
\ees
\ni
where ${\mbox{Li}}_2(z)$ is the polylogarithm function. As a result
the contribution of the longitudinal modes to the temperature
correction is
\bes
&&
\Delta_T^{||} F_{sl}(a)
=-\frac{\hbar cR}{8\pi a^3}
\sum\limits_{l=1}^{\infty}
\left\{
\frac{\pi}{2(lt)^3}\coth(\pi lt)-\frac{1}{(lt)^4}+
\frac{\pi^2}{2(lt)^2}\frac{1}{\sinh^2(\pi lt)}
\right.
\label{36} \\
&&\phantom{aaa}
+2\frac{\delta_0}{a}\left[\frac{\pi}{(lt)^3}\coth(\pi lt)
-\frac{3}{(lt)^4}+
\frac{\pi^2}{(lt)^2}\frac{1}{\sinh^2(\pi lt)}+
\frac{\pi^3}{lt}\frac{\coth(\pi lt)}{\sinh^2(\pi lt)}
\right]
\nn \\
&&\phantom{aaa}
-2\left(\frac{\delta_0}{a}\right)^2
\left[\frac{\pi}{(lt)^5}+
\frac{\pi^4}{\sinh^2(\pi lt)}
\left(1-3\coth^2(\pi lt)-\frac{\coth(\pi lt)}{\pi lt}
-\frac{2}{(\pi lt)^2}
\right)
\right.
\nn \\
&&\phantom{aaaaa}
\left.\left.
+\frac{6}{\pi(lt)^5}\left(
2\pi lt\ln\left(1-e^{-2\pi lt}\right)-
\frac{2\pi^2 (lt)^2}{e^{2\pi lt}-1}-
{\mbox{Li}}_2\left(e^{-2\pi lt}\right)\right)
\right]
\vphantom{\frac{\pi^3}{\sinh^2(\pi lt)}}
\right\}.
\nn
\ees
\ni
The contribution of the perpendicular modes is calculated 
simply as
\bes
&&
\Delta_T^{\bot} F_{sl}(a)
=-\frac{\hbar cR}{8\pi a^3}
\sum\limits_{l=1}^{\infty}
\left\{
\frac{\pi}{2(lt)^3}\coth(\pi lt)-\frac{1}{(lt)^4}+
\frac{\pi^2}{2(lt)^2}\frac{1}{\sinh^2(\pi lt)}
\right.
\nn \\
&&\phantom{aaa}
+2\frac{\delta_0}{a}
\left[\frac{\pi^3}{lt}\frac{\coth(\pi lt)}{\sinh^2(\pi lt)}
-\frac{1}{(lt)^4}
\right]
\label{37} \\
&&\phantom{aaa}
\left.
-2\left(\frac{\delta_0}{a}\right)^2\frac{\pi^3}{lt\sinh^2(\pi lt)}
\left[3\coth(\pi lt)+\pi lt -3\pi lt\coth^2(\pi lt)
\right]
\vphantom{\frac{\pi^3}{\sinh^2(\pi lt)}}
\right\}.
\nn
\ees

Putting together the contributions of both modes the total temperature
correction for the configuration of a lens above a plate is obtained:
\bes
&&
\Delta_T F_{sl}(a)
=-\frac{\hbar cR}{4\pi a^3}
\sum\limits_{l=1}^{\infty}
\left\{
\frac{\pi}{2(lt)^3}\coth(\pi lt)-\frac{1}{(lt)^4}+
\frac{\pi^2}{2(lt)^2}\frac{1}{\sinh^2(\pi lt)}
\right.
\label{38} \\
&&\phantom{aaa}
+\frac{\delta_0}{a}\left[\frac{\pi}{(lt)^3}\coth(\pi lt)
-\frac{4}{(lt)^4}+
\frac{\pi^2}{(lt)^2}\frac{1}{\sinh^2(\pi lt)}+
\frac{2\pi^3}{lt}\frac{\coth(\pi lt)}{\sinh^2(\pi lt)}
\right]
\nn \\
&&\phantom{aaa}
-\left(\frac{\delta_0}{a}\right)^2
\left[\frac{\pi}{(lt)^5}+
\frac{2\pi^4}{\sinh^2(\pi lt)}
\left(1-3\coth^2(\pi lt)+
\frac{\coth(\pi lt)}{\pi lt}-\frac{1}{(\pi lt)^2}\right)
\right.
\nn \\
&&\phantom{aaaaa}
\left.\left.
+\frac{6}{\pi(lt)^5}\left(
2\pi lt\ln\left(1-e^{-2\pi lt}\right)-
\frac{2\pi^2 (lt)^2}{e^{2\pi lt}-1}-
{\mbox{Li}}_2\left(e^{-2\pi lt}\right)\right)
\right]
\vphantom{\frac{\pi^3}{\sinh^2(\pi lt)}}
\right\}.
\nn
\ees
\ni
The terms of zeroth order in $\delta_0/a$ in the right-hand side of
(\ref{38}) coincide with the known result for an ideal metal.${}^{37}$
The coefficient of the first power in $\delta_0/a$ was 
already obtained in Ref.~23.

Now, let us consider the limiting cases of Eq.~(\ref{38}) corresponding
to low and high temperatures (small and large separations).
At low temperatures, $t\gg 1$,
and, preserving the largest of the exponentially small
contributions, one obtains from Eq.~(\ref{38})
\bes
&&
\Delta_T F_{sl}(a)\approx
-\frac{\hbar cR}{4\pi a^3}
\left\{\frac{\pi\zeta(3)}{2t^3}
-\frac{\pi^4}{90t^4}+\frac{2\pi^2}{t^2}e^{-2\pi t}
\right.
\label{39} \\
&&
\phantom{aaaaa}
\left.
+\frac{\delta_0}{a}\left[
\frac{\pi}{t^3}\zeta(3)-\frac{2\pi^4}{45t^4}
+\frac{8\pi^3}{t}e^{-2\pi t}\right]
-\left(\frac{\delta_0}{a}\right)^2
\left[\frac{\pi\zeta(5)}{t^5}-
16\pi^4e^{-2\pi t}\right]\right\}.
\nn
\ees
\ni
It is noticeable, that for the configuration of a lens above a plate the
second perturbative order in $\delta_0/a$  contains power-type
contributions in $t$, not only exponentially small ones. It is,
however, of order $t^{-5}$ in agreement with Ref.~24 where the
absence of temperature corrections of powers lower than $1/t^5$ was
proved for the perturbation orders $(\delta_0/a)^k$ with
$k=2,\,3,\,4,\,5,\,6$.

The limit of high temperatures can be obtained in the same way as
in Sec.3, i.e., starting from Eq.~(\ref{33}) and using Eq.~(\ref{26}).
The result is
\bes
&&
\Delta_T F_{sl}(a)\approx
-\frac{\hbar cR}{4\pi a^3}
\left\{\frac{\pi\zeta(3)}{2t}-\frac{\pi^4}{90}
\right.
\label{40} \\
&&
\phantom{aaaaa}
\left.
+\frac{\delta_0}{a}\left[
-\frac{\pi\zeta(3)}{t}+\frac{2\pi^4}{45}\right]
+\left(\frac{\delta_0}{a}\right)^2
\left[\frac{3\pi\zeta(3)}{t}-\frac{4\pi^4}{25}\right]
\right\}.
\nn
\ees

To determine the range of applicability of both asymptotic representations 
we compute the quantity
\beq
k_{sl}=\frac{\Delta_T F_{sl}(a)}{F_{sl}(a;T=0)},
\label{41}
\eeq
\ni
where $F_{sl}(a;T=0)$ is the force at zero temperature calculated
with account of finite conductivity.${}^{27-29}$
The value of $(1+k_{sl})$ has the meaning of a correction factor
to it.
The total Casimir force acting between real metals 
at nonzero temperature is given by
\beq
F_{sl}(a)=(1+k_{sl})F_{sl}(a;T=0).
\label{42}
\eeq

In Table 2 the values of $k_{sl}$ are presented  for aluminium
computed (i) by
Eq.~(\ref{38}) (second column), (ii) by the low-temperature asymptotic
(third column), and (iii) by the high-temperature asymptotic (fourth
column).
In analogy with the case of two parallel plates, the asymptotics of
low and high temperatures work good at separations $a\leq 2\,\mu$m
and $a\geq 6\,\mu$m, respectively. The data of column 2 are practically
the same in first- and second-order 
perturbation theory. They coincide also 
with the results of numerical computations by the use of Eq.~(\ref{33}).

The comparative contribution of the longitudinal and perpendicular
modes to the temperature correction to the Casimir force
(see Eqs.~(\ref{36}), (\ref{37})) is different in comparision 
with the case
of two plates. Here,
$\Delta_{T}^{||}F_{sl}/\Delta_{T}^{\bot}F_{sl}=1.64$
at $a=0.1\,\mu$m, 1.21 at $a=0.3\,\mu$m, and 1.13 at $a=0.5\,\mu$m.
This ratio decreases slowly to the value 1.006 at $a=10\,\mu$m.
Thus, the contributions of both modes are approximately equal to 
each other at all separations (for two plates the contribution
of the longitudinal mode significantly dominates at smallest
separations). This is explained by the presence of the term
$\sim t^{-3}$ in the zeroth order contribution to $\Delta_T F_{sl}$
(in the case of two plates the zeroth order contribution is
$\sim t^{-4}$). 

\section{Casimir force with account of roughness}

The above Eqs.~(\ref{28}), (\ref{41}) give us the Casimir force
computed at nonzero temperature with account of finite conductivity
of the boundary metal. Except for the finite value of conductivity,
real metallic boundaries are characterized also by some surface
roughness. According to the results of Ref.~35 obtained at zero
temperature, for a wide range of surface roughness it can be taken
into account by some kind of geometrical approach using the averaging
of the Casimir force over the rough surface. Here, we generalize this
approach for the case of nonzero temperature.

Let two large metallic plates of dimension $L\times L$ be covered by
small roughness. Then the distance between two points of the boundary
surfaces of different plates with the coordinates ($x,y$) can be 
expressed as
\beq
a(x,y)=a_0+f(x,y).
\label{43}
\eeq
\ni
Here, $f(x,y)$ is simply expressed  by the functions
describing roughness on both surfaces. The mean distance between
the plates $a_0$ is defined in such a way that
\beq
\int\limits_{-L}^{L}dx
\int\limits_{-L}^{L}dy
f(x,y)=0.
\label{44}
\eeq

As a result, the Casimir force taking all real conditions into
account (i.e. nonzero temperature, finite conductivity of a metal
and surface roughness) is given as
\beq
F_{ss}^r(a_0)=\frac{1}{L^2}
\int\limits_{-L}^{L}dx
\int\limits_{-L}^{L}dy
F_{ss}[a(x,y)].
\label{45}
\eeq
\ni
Remind that $F_{ss}$ here is given by Eq.~(\ref{28}) with a change of
separation distance $a$ for the one defined in Eq.~(\ref{43}).
The analogical result can be obtained for the configuration of a sphere
above a plate by the use of the proximity force theorem.${}^{46}$
As is shown in Ref.~37, Eq.~(\ref{45}) gives the same results as 
the more fundamental methods for accounting roughness, e.g., based on
the specific forms of interatomic potentials or Green's
function method.

Finite conductivity and temperature corrections to the Casimir force
were computed in Secs.3,\,4 in the whole distance range
$0.1\,\mu\mbox{m}\leq a\leq 10\,\mu$m. Surface roughness makes the
most important contribution to the Casimir force for separation
distances  $a\leq 1\,\mu$m. For such distances (and also for
$1\,\mu\mbox{m}< a< 3\,\mu$m) the asymptotic of low temperatures
of Eq.~(\ref{25}) can be substituted into Eq.~(\ref{45}) to calculate
the Casimir force under real conditions. As to the transition region
$3\,\mu\mbox{m}\leq a\leq 5\,\mu$m, the more exact Eq.~(\ref{24})
should be used there (at larger separations roughness corrections
are negligible).

\section{Conclusions and discussion}

As was argued above, the dielectric function as it results from the
plasma model can be reliably used to calculate the
Casimir force acting between real metals at nonzero temperature.
The scattering theory underlying the Lifshitz formula is well defined
in the case of the plasma model and its application is straightforward.
No additional prescriptions are needed like those formulated in
Ref.~22 for ideal metal or in Ref.~41 for real metals described by
the Drude dielectric function.

We calculated the temperature correction to the Casimir force
between real metals in the configuration of two parallel plates (two
semispaces) and for a lens (sphere) above a plate. The analytical
expressions for these corrections were obtained which are exact with
respect to the temperature but perturbative with respect to the
effects of finite conductivity. These effects were taken into
account up to the second order in a small parameter having the meaning
of the relative penetration depth of electromagnetic zero-point
oscillations into the metal. The asymptotics of the obtained
expressions were presented at both low and high temperatures 
relative to $k_BT_{eff}=\hbar c/(2a)$.
The asymptotical formulas are in good agreement with 
the exact ones except of a narrow transition region between the cases 
of low and high temperatures. The scopes of these regions are
determined. In a wide separation range from 0.1$\,\mu$m till
10$\,\mu$m the obtained analytical results are in perfect accordance with
the results of numerical computations performed earlier. The comparative
contributions of the longitudinal and perpendicular modes to the
temperature correction were determined. Modification of the obtained
results taking the surface roughness into account was given.
This together permits to evaluate the Casimir force under the
influence of real conditions which include
nonzero temperature, finite conductivity of the boundary
metal and surface roughness with a precision of several percent.

The above results are topical ones for the precision measurements of the
Casimir force. In view of fundamental and technological applications
of the Casimir effect mentioned in Introduction, there is a great
theoretical challenge to account for real experimental conditions.
In fact, a theory is required which makes it possible to calculate
the Casimir force with a precision of being
better than one percent.
For this purpose one should especially examine the optical properties
of the test bodies in use, their surface roughness and take into
account the spatial dispersion (in the case when there are thin layers 
covering the test bodies). The more precise future
theory should take into account also the effects of dissipation which 
are neglected in the presently used Lifshitz formula.

\section*{Acknowledgements}

The authors are grateful to M.~Bordag for several helpful discussions.
G.L.K. and V.M.M. are indebted to the Center of Theoretical Studies and
the Institute for Theoretical
Physics, Leipzig University, for kind hospitality. Their work was
supported by the Saxonian Ministry of Science and Fine Arts (Germany)
and by CNPq (Brazil).
 
\newpage
\begin{table}[h]
\caption{The values of $k_{ss}$ computed by the use of
Eq.~(\ref{24}) in comparison with the asymptotic values at low
$[$Eq.~(\ref{25})$]$ and high  $[$Eq.~(\ref{27})$]$ 
temperatures.}
\vspace*{1.5cm}
\begin{tabular}{cccc}
{\normalsize{$a$ }}& {\normalsize{$k_{ss}$ computed }}
 & {\normalsize{$k_{ss}$ at low }}
& {\normalsize{$k_{ss}$ at high}}\\
{\normalsize{$(\mu\mbox{m})$ }} &
{\normalsize{by Eq.~(\ref{24})}} 
& {\normalsize{temperatures}}
& {\normalsize{temperatures }}\\
&&& \\
0.1 & 6.58$\times 10^{-6}$ & 6.58$\times 10^{-6}$ & \\
0.3 & 5.48$\times 10^{-5}$ & 5.48$\times 10^{-5}$ &  \\
0.5 & 2.12$\times 10^{-4}$ & 2.12$\times 10^{-4}$ &  \\
0.7 & 6.05$\times 10^{-4}$ & 6.05$\times 10^{-4}$ &  \\
0.9 & 1.42$\times 10^{-3}$ & 1.42$\times 10^{-3}$ &  \\
2.0 & 2.74$\times 10^{-2}$ & 2.76$\times 10^{-2}$ &  \\
3.0 & 0.1228 & 0.1339 &  \\
4.0 & 0.3100 & 0.4159  & 0.2302 \\
5.0 & 0.5630 & & 0.5349 \\
6.0 & 0.8487 & & 0.8400 \\
7.0 & 1.147 & & 1.144 \\
8.0 & 1.450 & & 1.449\\
10 & 2.059 & & 2.059 
\end{tabular}
\end{table}
\newpage
\begin{table}[h]
\caption{The values of $k_{sl}$ computed by the use of
Eq.~(\ref{38}) in comparison with the asymptotic values at low
$[$Eq.~(\ref{39})$]$ and high  $[$Eq.~(\ref{40})$]$ 
temperatures.}
\vspace*{1.5cm}
\begin{tabular}{cccc}
{\normalsize{$a$ }}& {\normalsize{$k_{sl}$ computed }}
 & {\normalsize{$k_{sl}$ at low }}
& {\normalsize{$k_{sl}$ at high}}\\
{\normalsize{$(\mu\mbox{m})$ }} &
{\normalsize{by Eq.~(\ref{38})}} 
& {\normalsize{temperatures}}
& {\normalsize{temperatures }}\\
&&& \\
0.1 & 6.69$\times 10^{-5}$ & 6.70$\times 10^{-5}$ & \\
0.3 & 1.08$\times 10^{-3}$ & 1.08$\times 10^{-3}$ &  \\
0.5 & 4.33$\times 10^{-3}$ & 4.33$\times 10^{-3}$ &  \\
0.7 & 1.09$\times 10^{-2}$ & 1.09$\times 10^{-2}$ &  \\
0.9 & 2.181$\times 10^{-2}$ & 2.181$\times 10^{-2}$ &  \\
2.0 & 0.1829 & 0.1828 &  \\
3.0 & 0.4810 & 0.4762 & 0.3857 \\
4.0 & 0.8736 & 0.8113  & 0.8428 \\
5.0 & 1.309 & & 1.300 \\
6.0 & 1.759 & & 1.757 \\
7.0 & 2.215 & & 2.214 \\
8.0 & 2.671 & & 2.671\\
10 & 3.585 & & 3.585 
\end{tabular}
\end{table}
\end{document}